%% file: m80_chemo_dynamics.tex
\documentclass[a4paper,fleqn,usenatbib,useAMS]{mnras}

\usepackage{newtxtext,newtxmath}

\usepackage[T1]{fontenc}
\usepackage{ae,aecompl}

\usepackage{graphicx}	
\usepackage{amsmath}	
\usepackage{amssymb}	
\usepackage[dvipsnames]{xcolor}

\usepackage{etoolbox}
\makeatletter
\patchcmd\@combinedblfloats{\box\@outputbox}{\unvbox\@outputbox}{}{%
    \errmessage{\noexpand\@combinedblfloats could not be patched}%
}%
\makeatother

\title[Kinematics of multiple populations in M80]{The peculiar kinematics of the multiple populations in the globular cluster Messier~80 (NGC~6093)}

\author[S. Kamann et al.]{%
S. Kamann,$^{1}$\thanks{E-mail: s.kamann@ljmu.ac.uk}
E. Dalessandro,$^{2}$
N. Bastian,$^{1}$
J. Brinchmann,$^{3,4}$
M. den Brok,$^{5}$
\newauthor
S. Dreizler,$^{6}$
B. Giesers,$^{6}$
F. G\"ottgens,$^{6}$
T.-O. Husser,$^{6}$
D. Krajnovi\'c,$^{5}$
\newauthor
G. van de Ven,$^{7}$
L.~L. Watkins,$^{7,8}$
L. Wisotzki,$^{5}$
\\
$^{1}$Astrophysics Research Institute, Liverpool John Moores University, 146 Brownlow Hill, Liverpool L3 5RF, UK\\
$^{2}$INAF -- Astrophysics and Space Science Observatory Bologna, Via Gobetti 93/3, I-40129 Bologna, Italy\\ 
$^{3}$Instituto de Astrof{\'\i}sica e Ci{\^e}ncias do Espa\c co, Universidade do Porto, CAUP, Rua das Estrelas, PT4150-762 Porto, Portugal\\
$^{4}$Leiden Observatory, Leiden University, P.O. Box 9513, 2300 RA, Leiden, The Netherlands\\
$^{5}$Leibniz-Institute for Astrophysics, An der Sternwarte 16, 14482 Potsdam, Germany\\
$^{6}$Institute for Astrophysics, Georg-August-Universit\"at G\"ottingen, Friedrich-Hund-Platz 1, 37077 G\"ottingen, Germany\\
$^{7}$Department of Astrophysics, University of Vienna, T{\"u}rkenschanzstra{\ss}e 17, 1180 Vienna, Austria \\
$^{8}$ESO, European Southern Observatory, Karl-Schwarzschild Str. 2, 85748 Garching bei M\"unchen, Germany
}

\date{Accepted XXX. Received YYY; in original form ZZZ}

\pubyear{2019}

\begin{document}
\label{firstpage}
\pagerange{\pageref{firstpage}--\pageref{lastpage}}
\maketitle

\begin{abstract}
We combine MUSE spectroscopy and {\it Hubble Space Telescope} ultraviolet (UV) photometry to perform a study of the chemistry and dynamics of the Galactic globular cluster Messier~80 (M80, NGC~6093). Previous studies have revealed three stellar populations that not only vary in their light-element abundances, but also in their radial distributions, with concentration decreasing with increasing nitrogen enrichment. This remarkable trend, which sets M80 apart from the other Galactic globular clusters, points towards a complex formation and evolutionary history. To better understand how M80 formed and evolved, revealing its internal kinematics is key. We find that the most N-enriched population rotates faster than the other two populations at a $2\sigma$ confidence level. While our data further suggest that the intermediate population shows the least amount of rotation, this trend is rather marginal ($1-2\sigma$). Using axisymmetric Jeans models, we show that these findings can be explained from the radial distributions of the populations if they possess different angular momenta. Our findings suggest that the populations formed with primordial kinematical differences.

\end{abstract}

\begin{keywords}
Galaxy: globular clusters: individual: M80 -- stars: kinematics and dynamics -- stars: abundances
\end{keywords}



\section{Introduction}



The occurrence of multiple populations in massive star clusters is still an unsolved puzzle. Such populations are characterized by subtle differences in their light-element abundances \citep[like, C, N, Na, O, e.g.][]{2009A&A...505..117C} and manifest as slightly different tracks across color magnitude diagrams when appropriate near-UV filter combinations are used \citep[e.g.][]{2017MNRAS.464.3636M}. They seem to be omnipresent in clusters more massive than $\sim10^4-10^5\,{\rm M_\odot}$ and older than $2~{\rm Gyr}$ \citep{2018MNRAS.473.2688M}. While each cluster appears to have its own abundance fingerprint, certain characteristics apply to all clusters studied to date. In particular, each cluster harbours a \emph{primordial} population characterized by an abundance pattern similar to that of field stars with comparable metallicity, alongside one or more populations showing enrichment in some elements (like N or Na) and depletion in others (such as C or O) compared to the primordial population. Various scenarios have been advocated to explain the presence of these \emph{N-enriched} populations. However, none of them seems to be able to explain the wealth of observational data that is available \citep[see][for a review]{2018ARA&A..56...83B}.

Despite the short half-mass relaxation times of most clusters, possible kinematic signatures imprinted at the time of formation of the populations could still be observable today \citep[e.g.][]{2015MNRAS.450.1164H}. So investigating the kinematics of the different populations is a promising way to make progress. Albeit challenging, a number of observational studies in this direction have been performed. Proper motion studies found evidence for higher radial anisotropies of the N-enriched populations in some clusters, such as 47Tuc \citep{2013ApJ...771L..15R,2018MNRAS.479.5005M}, NGC~2808 \citep{2015ApJ...810L..13B}, $\omega$~Cen \citep{2018ApJ...853...86B}, and possibly also NGC~362 \citep{2018ApJ...861...99L}. This behaviour can be explained if the N-enriched stars are more centrally concentrated at the time of formation and are then scattered on radial (wider) orbits as relaxation proceeds. An enhanced concentration of N-enriched stars appears to be still present in the majority of clusters studied to date \citep[see][and references therein]{2019ApJ...884L..24D}. Indeed, complete mixing between the populations is expected only in clusters that have experienced significant mass loss and are in an advanced dynamical state \citep[see for example the case of NGC~6362,][]{2014ApJ...791L...4D,2015MNRAS.454.2166M,2013MNRAS.429.1913V}.

\begin{table*}
	\centering
	\caption{Summary of MUSE observations of M80.}
	\label{tab:observations}
	\begin{tabular}{cccccc} 
		\hline
		Pointing & RA & Dec & Obs. date & Seeing & Exp. time \\
		\hline
1 & 16:17:00.78 & -22:58:56.4 &  2015-05-11 07:56:56 & 0.62\arcsec & $3\times200\,{\rm s}$ \\
 & & & 2017-04-23 09:10:55 & 0.66\arcsec & $3\times200\,{\rm s}$ \\ \hline
2 & 16:17:00.78 & -22:58:11.4 & 2015-05-11 08:12:39 & 0.62\arcsec & $3\times200\,{\rm s}$ \\
 & & & 2017-04-23 09:29:43 & 0.66\arcsec & $3\times200\,{\rm s}$ \\
 & & & 2017-02-01 09:11:41 & 0.52\arcsec & $3\times200\,{\rm s}$ \\ \hline
3 & 16:17:04.04 & -22:58:56.4 & 2015-05-11 08:42:29 & 0.64\arcsec & $3\times200\,{\rm s}$ \\
 & & & 2015-05-11 08:58:28 & 0.74\arcsec & $3\times200\,{\rm s}$ \\
 & & & 2017-04-26 04:22:07 & 0.92\arcsec & $3\times200\,{\rm s}$ \\ \hline
4 & 16:17:04.04 & -22:58:11.4 & 2015-05-11 09:14:19 & 0.64\arcsec & $3\times200\,{\rm s}$ \\
 & & & 2017-04-26 04:37:08 & 0.82\arcsec & $2\times200\,{\rm s}$ \\
\hline
	\end{tabular}
\end{table*}

However, virtually all scenarios put forward to explain the presence of multiple populations predict that N-enriched stars are centrally concentrated at formation. Hence, the discriminatory power of differences in the anisotropy patterns is limited \citep{2015MNRAS.450.1164H}. Rotation may be a more powerful diagnostic in this respect. It has been detected in a significant number of clusters \citep[e.g][]{2012A&A...538A..18B,2014ApJ...787L..26F} and recent findings on a correlation between angular momentum and relaxation time \citep{2018MNRAS.473.5591K,2018MNRAS.481.2125B,2019MNRAS.485.1460S} suggest that it played a crucial role during the formation of the clusters. Furthermore, \citet{2015MNRAS.450.1164H} showed that scenarios involving multiple epochs of star formation predict that later generation(s) should form with higher rotation velocities because angular momentum must be conserved during the infall of gas expelled by the first generation towards the cluster centre \citep[see also][]{2010ApJ...724L..99B,2013ApJ...779...85M}. On the other hand, the opposite is expected in scenarios where the different populations form simultaneously. If supermassive stars play a role in the formation of multiple populations, as recently advocated by \citet{2018MNRAS.478.2461G}, the N-enriched population is expected to pick up the angular momentum of said stars, causing it to rotate slowly, irrespective of the rotation speed of the primordial population. The spin axes of the two populations could also be misaligned, with the N-enriched population counter-rotating or corotating relative to the primordial one.

Only a handful of observational studies have looked at the rotation of different populations, and found varying results. While no differences were found in 47~Tuc \citep{2018MNRAS.479.5005M}, NGC~6362 \citep{2018ApJ...864...33D}, or NGC~6352 \citep{2019arXiv190202787L}, \citet{2018ApJ...853...86B} found rotational differences among the populations of the complex cluster $\omega$~Cen, in the sense that main sequence stars with enhanced helium and iron abundances rotate slower. On the other hand, \citet{2017MNRAS.465.3515C} reported enhanced rotation for the extremely N-enriched population in NGC~6205 (M13). Surprisingly, NGC~6205 also appears to be in a late dynamical stage, given that \citet{2018MNRAS.474.4438S} found the populations are almost completely mixed\footnote{Note that \citet{2018MNRAS.474.4438S} only divided their sample into primordial and enriched stars, whereas \citet{2017MNRAS.465.3515C} further divided the latter into an intermediate and an extreme population. \citet{2012ApJ...754L..38J} find the extreme population to be more centrally concentrated than the other two.}. This shows the need for further studies of clusters in all evolutionary stages in order to use rotation as an ingredient in solving the puzzle of multiple populations.

In this paper, we study the globular cluster NGC~6093 (M80). \citet{2018ApJ...859...15D} recently found the three detected populations to be unusually distributed, with the primordial population being more centrally concentrated than the intermediate (in terms of N-enrichment) population, which in turn is more centrally concentrated than the extreme population. NGC~6093 is considered to be dynamically old \citep{2012Natur.492..393F}, in which case the different concentrations are unlikely to be a relic from the formation of the cluster. Instead, \citet{2018ApJ...859...15D} suggested that helium variations of $\Delta Y\sim0.05-0.06$ cause the N-enriched stars to be less massive, so that the different radial distributions can be explained by mass segregation. In this work, we study the kinematics of the populations and investigate if they hold further clues on the dynamical evolution of NGC~6093. With this aim, we combine the photometry of \citet{2018ApJ...859...15D} with MUSE \citep{2010SPIE.7735E..08B} integral field spectroscopy.

This paper is organized as follows. The MUSE data are introduced in Sect.~\ref{sec:data} and matched to the photometry in Sect.~\ref{sec:analysis}. In Sect.~\ref{sec:morphology}, we investigate the morphology of NGC~6093 before turning to the cluster kinematics in Sections~\ref{sec:kinematics} and \ref{sec:models}. We conclude in Sect.~\ref{sec:conclusions}.

\section{Spectroscopic data}
\label{sec:data}


The spectroscopy used in this study was obtained as part of the MUSE guaranteed time observations, in the programme ``A stellar census in globular clusters with MUSE''. A detailed summary of the programme, the data reduction and their analysis is provided in \citet{2018MNRAS.473.5591K}. Here, we restrict ourselves to a brief overview of the main aspects.

In Table~\ref{tab:observations}, we list all the observations that have been used in the current study. Four pointings in M80 have been repeatedly observed. They cover the central region of the cluster out to a distance of $\sim1\arcmin$, i.e., beyond its half-light radius of $r_{\rm h}=36\arcsec$ \citep{2018MNRAS.478.1520B}. Each observation listed in Table~\ref{tab:observations} consisted of three exposures, offset by $90\,{\rm degrees}$ in the derotator angle. The individual exposures were processed with the standard MUSE pipeline \citep{2012SPIE.8451E..0BW,2014ASPC..485..451W}, which was also used to create a combined data cube for each set of three exposures afterwards. One failed exposure of pointing 4 was excluded from the combination process. Although the data were taken before the installation of the new adaptive optics system, the image quality is very good, with an average seeing of $0.7\arcsec$ as measured on whitelight images created from the combined cubes.

The spectra of the individual stars cover a wide wavelength range ($480\,{\rm nm} < \lambda < 930\,\rm{nm}$) at low to medium spectral resolution ($R\sim1\,700-3\,500$). They were extracted from the data cubes using the \textsc{PampelMuse} software described in \citet{2013A&A...549A..71K}. The input photometry required to extract the spectra was obtained as part of the HST/ACS survey of Galactic globular clusters \citep{2007AJ....133.1658S,2008AJ....135.2055A}.

After extraction, the spectra are subjected to a number of analyses. Most notably for the current paper, the radial velocity of every spectrum is determined in a two-step procedure. An initial value is obtained via cross-correlation against a set of template spectra. The best-matching template is selected via the widely-used $r_{\rm cc}$ parameter \citep{1979AJ.....84.1511T} and the velocity $v_{\rm LOS, cc}$ it yielded is used as input for the following full-spectrum fit, which is explained in detail in \citet{2016A&A...588A.148H}. In brief, each extracted spectrum is fitted against the synthetic library presented in \citet{2013A&A...553A...6H} to derive a metallicity ${\rm [M/H]}$, an effective temperature $T_{\rm eff}$, and a radial velocity $\Tilde{v}_{\rm LOS,\,fit}$. The initial guesses for ${\rm [M/H]}$ and $T_{\rm eff}$ are obtained from the catalog of \citet{1996AJ....112.1487H} and a comparison between the input photometry with an isochrone from the database of \citet{2017ApJ...835...77M}, respectively. The latter also yields a surface gravity $\log g$. Given the challenges involved in measuring $\log g$ spectroscopically at the resolution of the MUSE data, the surface gravity is currently held constant at the value obtained from the isochrone comparison.

As outlined in \citet{2016A&A...588A.149K}, the telluric absorption components that are included in the full-spectrum fit allow us to validate the accuracy of the wavelength solution. For each observation listed in Table~\ref{tab:observations}, we used all spectra extracted with ${\rm S/N} > 20$\footnote{Measured per pixel and averaged over the full spectrum.} and determined the mean velocity ${\overline v}_{\rm LOS, tell}$ of the telluric components relative to the expected barycentric velocity. The result was subtracted from all stellar velocities derived from the observation. This yielded the final velocity $v_{\rm LOS, fit} = \Tilde{v}_{\rm LOS,\,fit} - {\overline v}_{\rm LOS, tell}$ for each extracted spectrum. The measurement uncertainties tailored to the $\Tilde{v}_{\rm LOS,\,fit}$ values were corrected for residual errors in the wavelength calibration by adding in quadrature the standard deviation of the telluric components across all spectra extracted with ${\rm S/N} > 20$.

\section{Kinematic samples}
\label{sec:analysis}

\begin{figure}
 \includegraphics[width=\columnwidth]{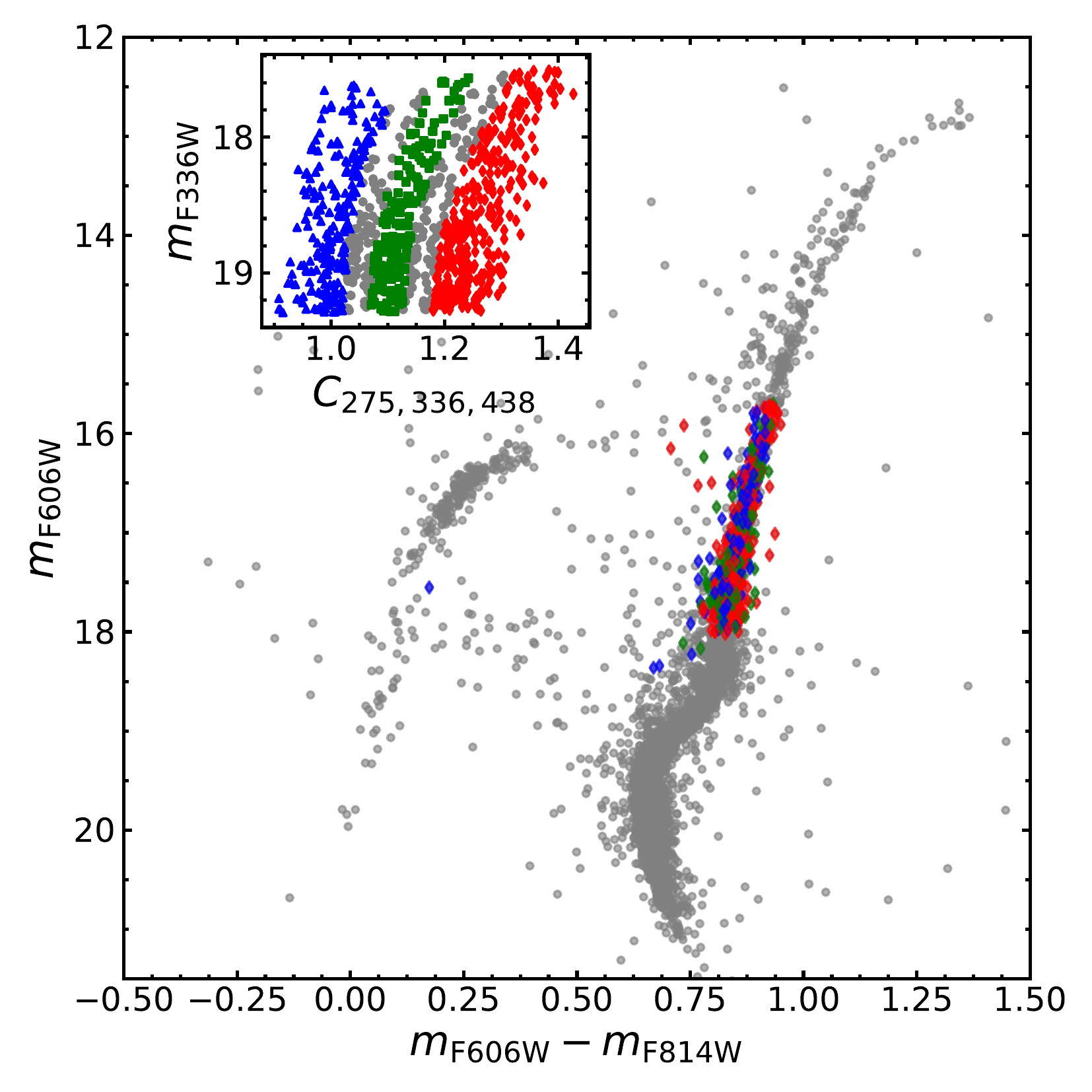}
 \caption{Color-magnitude diagram (CMD) of the stars in NGC~6093 for which radial velocities from MUSE are available. The main panel shows the optical CMD of the entire sample. The stars that we could identify as members of the \emph{primordial} (red), \emph{intermediate} (green), and \emph{extreme} (blue) populations are colour-coded. Their location in an UV pseudo-CMD is shown in the inset panel. Stars with uncertain population membership in the displayed magnitude range are shown in grey.}
 \label{fig:cmd}
\end{figure}

From our full sample of radial velocities derived from the MUSE data, we excluded those stemming from spectra extracted with ${\rm S/N} < 7$, which we found to be the limit for reliable velocity determinations in \citet{2018MNRAS.473.5591K}. Further, we imposed that the velocities derived from the cross correlation and the full spectrum fit, $v_{\rm LOS, cc}$ and $v_{\rm LOS, fit}$, were offset by no more than $3\times$ the propagated uncertainty of their difference, and that the cross correlation yielded a reliable signal, indicated by $r_{\rm cc} > 4$ \citep[cf.][]{2018MNRAS.473.5591K}. Finally, we also excluded velocities derived from spectra for which the recovered F606W magnitude deviated strongly (by more than $2\times$ the standard deviation obtained for stars of comparable brightness) from the one available in the input catalog. In such cases, it is likely that the extracted spectrum is contaminated by flux originating from a nearby (brighter) star. The cuts left us with a sample of $12\,652$ spectra of $6\,284$ stars. In Fig.~\ref{fig:cmd}, we show the positions of the remaining stars in an optical colour-magnitude diagram of NGC~6093.

We derived stellar velocities by combining the $v_{\rm LOS, fit}$ measurements from the spectra belonging to the individual stars and propagating their uncertainties. As in \citet{2018MNRAS.473.5591K}, we used the scatter of the $v_{\rm LOS, fit}$ measurements per star to calibrate the final uncertainties. As a consequence of the large range in stellar magnitudes (cf. Fig.~\ref{fig:cmd}), spectra are extracted over a large range in signal-to-noise, resulting in a broad distribution of radial velocity uncertainties. The 16th, 50th, and 84th percentiles of said distribution are $2.3\,{\rm km\,s^{-1}}$, $6.9\,{\rm km\,s^{-1}}$, and $10.7\,{\rm km\,s^{-1}}$, respectively.

The availability of multiple measurements also allows us to identify stars with varying radial velocities. Each star with multiple measurements was assigned a probability of being variable. However, due to the limited temporal coverage of our observations (cf. Table~\ref{tab:observations}), our dataset does not currently allow us to make firm conclusions about binarity. In the present study, we removed stars with variability probabilities $>80\%$ from all analyses of the cluster kinematics.

We matched the MUSE sample to the HST photometry by \citet{2018ApJ...859...15D} to distinguish among different stellar sub-populations with different light-element abundance in our radial velocity sample. 
\citet{2018ApJ...859...15D} identified three different sub-populations along the RGB (for $17.7<m_{F336W}<19.4$) in the verticalised $\Delta_C(m_{F275W}-m_{F336W})-(m_{F336W}-m_{F438W})$ pseudo-color diagram, which were labeled as first-generation stars (FG), intermediate second-generation stars (SG$_{\rm INT}$), and extreme second-generation stars (SG$_{\rm EXT}$) based on their colours. As the origin of the sub-populations is still debated and their formation may not happen in a temporal sequence (as suggested by the word \emph{generations}), we will simply refer to the three groups as primordial (=FG), intermediate (=SG$_{\rm INT}$), and extreme (=SG$_{\rm EXT}$) populations throughout the paper.

After accounting for a small global offset between the \textit{HST} catalogues, we considered each source \textit{from the final MUSE radial velocity sample} as matched if it had a counterpart within $0.01\arcsec$ in the \citet{2018ApJ...859...15D} data. In total, we were able to recover 733 out of the 943 RGB stars used by \citet{2018ApJ...859...15D} in the MUSE data. Their location in an UV pseudo-CMD is shown in the inset panel in Fig.~\ref{fig:cmd}. The stars that we could not recover are mainly outside the MUSE field of view or located within $\sim10\arcsec$ from the cluster centre, where the stellar density is so high that the MUSE sample becomes incomplete at the bottom of the red giant branch.

Following \citet{2018ApJ...859...15D}, we only assigned stars to a certain population if their membership probability of belonging to said population exceeded $P=85\%$. This is the case for 714 out of the 943 RGB stars, where 559 out of the 714 stars are also included in the MUSE sample. We recovered 271 of 325 stars from the primordial population, 126 of 162 stars from the intermediate one, and 162 of 227 stars from the extreme population. In both panels of Fig.~\ref{fig:cmd}, we highlight the members of the three populations that we were able to recover in the MUSE data. As the stars are found towards the bright end of the full MUSE sample, their radial velocities are typically measured more accurately compared to the full sample. The 16th, 50th, and 84th percentiles of the radial velocity uncertainty distribution for the stars with population tags are $1.3\,{\rm km\,s^{-1}}$, $1.8\,{\rm km\,s^{-1}}$, and $2.8\,{\rm km\,s^{-1}}$, respectively.

We complemented the MUSE kinematical data with the radial velocities compiled by \citet{2018MNRAS.478.1520B}. Their sample includes $232$ stars in NGC~6093, measured to an accuracy of typically $1.3\,{\rm km\,s^{-1}}$ and mainly located outside the half-light radius of the cluster. We did not try to assign them to any of the stellar populations, however they will be very valuable in constraining the dynamical models (cf. Sect.~\ref{sec:models}) in the outskirts of the cluster. A total of $13$ stars overlap between the two samples. Their velocity measurements are in good agreement, with an average offset of $-0.8\,{\rm km\,s^{-1}}$ between the MUSE and literature data and a $\chi^2$ of $11.7$ summed over the $13$ stars.

Prior to any analysis, we determined the mean velocity of the cluster using both the MUSE and the literature sample, yielding $(10.0\pm0.1)\,{\rm km\,s^{-1}}$ and $(10.3\pm0.3)\,{\rm km\,s^{-1}}$, respectively. These values were subtracted from the samples before they were combined and the systemic velocity of the cluster was fixed to $0\,{\rm km\,s^{-1}}$ during the analyses. In addition, we corrected all velocities for the effect of perspective rotation, using the approach of \citet{2006A&A...445..513V} and the systemic proper motion of NGC~6093 determined by \citet[$\mu_{\alpha*}=-2.95\,{\rm mas\,yr^{-1}}$, $\mu_{\delta}=-5.56\,{\rm mas\,yr^{-1}}$]{2018A&A...616A..12G}.

\section{Cluster morphology}
\label{sec:morphology}

\subsection{Radial density profiles}

\begin{figure}
 \includegraphics[width=\columnwidth]{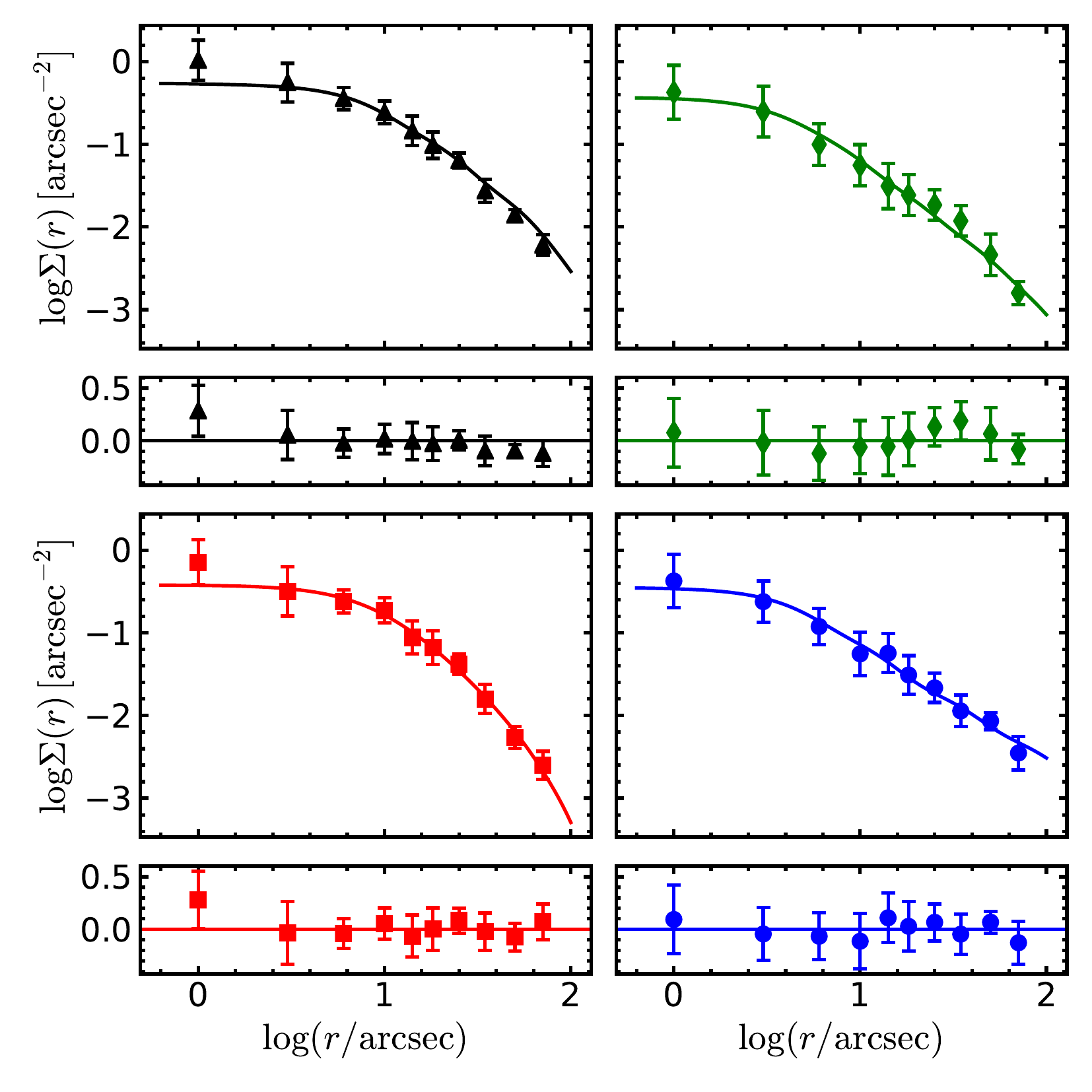}
 \caption{Surface density profiles of different stellar populations in NGC~6093: entire cluster (\textit{top left}), primordial population (\textit{bottom left}), intermediate population (\textit{top right}), and extreme population (\textit{bottom right}). In each major panel, we show the actual star counts and the best-fit MGE profile. Minor panels show the residuals after subtracting the MGE fits from the data.}
 \label{fig:surface_brightness}
\end{figure}

We derived the observed surface density profiles of the three sub-populations selected by \citet{2018ApJ...859...15D} as well as of the entire RGB population in the same magnitude range that was used to identify the sub-populations in Sect.~\ref{sec:analysis}.
For each sub-population, we divided the FOV into 10 concentric annuli centered on the centre of gravity obtained by \citet{2018ApJ...859...15D}, ($\alpha=16^h17^m2^s.481, \delta=-22^{\circ}58'34''.098$). Each annulus was split into four sub-sectors, with the exception of the most external one which was split into only two sub-sectors. Number counts of the stars in the \citet{2018ApJ...859...15D} sample were then calculated in each sub-sector and the corresponding densities were obtained by dividing them by the sampled area. The stellar density of each annulus was defined as the average of the sub-sector densities, and its standard deviation was computed from the variance among the sub-sectors. In light of the results of \citet{2018ApJ...864...33D}, who used artificial star tests to infer that their star counts are complete at a $>95\%$ level, no incompleteness corrections were applied. The resulting surface density profiles for the four populations are shown in Fig~\ref{fig:surface_brightness}.

\begin{table*}
	\centering
	\caption{Parameters of the MGE models fitted to the different stellar populations in NGC~6093. For each Gaussian component $k$, the standard deviation $\sigma_{k}$ and the logarithm of the central number density $\Sigma_{k,\,{\rm 0}}$ are provided.}
	\label{tab:mge_parameters}
    \include{ngc6093_mge_parameters}
\end{table*}

As the photometry of \citet{2018ApJ...859...15D} is limited to the central region of the cluster, we complemented our global number density profile with the \emph{Gaia} data recently presented by \citet{2019MNRAS.485.4906D}. After accounting for a vertical offset, the two profiles were stitched together. Note that the number densities underlying both profiles are dominated by red giant stars, so their combination is not affected by processes such as mass segregation. The resulting combined profile was fitted with a one-dimensional Multi-Gaussian Expansion \citep[MGE,][]{1994A&A...285..723E}, using the code of \citet{2002MNRAS.333..400C}. Such an MGE representation is required for the dynamical modeling performed in Sect.~\ref{sec:models}, hence we repeated the fit for the three sub-populations. In the absence of population tags for the \emph{Gaia} data, we assumed that the outskirts of the sub-population profiles follow isotropic single-mass \citet{1966AJ.....71...64K} models that we fitted to the profiles derived by \citet{2018ApJ...859...15D}. We obtained concentrations and half-light radii of ($c=1.32$, $r_{\rm h}=37\arcsec$), ($c=1.95$, $r_{\rm h}=44\arcsec$), and ($c=2.42$, $r_{\rm h}=188\arcsec$) for the primordial, intermediate, and extreme populations, respectively. To represent the profiles as MGEs, six Gaussian components were required. The final MGE models are included in Fig.~\ref{fig:surface_brightness} and their parameters are summarized in Table~\ref{tab:mge_parameters}.

In addition, we also converted the global number density profile into a projected luminosity density profile. The latter will be used in Sect.~\ref{sec:models} to infer the mass-to-light ratio of the cluster. Under the assumption that the average luminosity per star does not change with radius, the conversion is just a multiplication with a constant factor. We determined this factor by enforcing that after integrating over the profile and correcting for an extinction of $A_{\rm V}=0.56$ \citep[assuming $A_{\rm V}=3.1*E_{\rm B-V}$ and using $E_{\rm B-V}= 0.18$,][]{1996AJ....112.1487H}, we obtained an apparent cluster magnitude of $V=7.33$ \citep{1996AJ....112.1487H}.

\subsection{Cluster elongation}

\begin{figure}
 \includegraphics[width=\columnwidth]{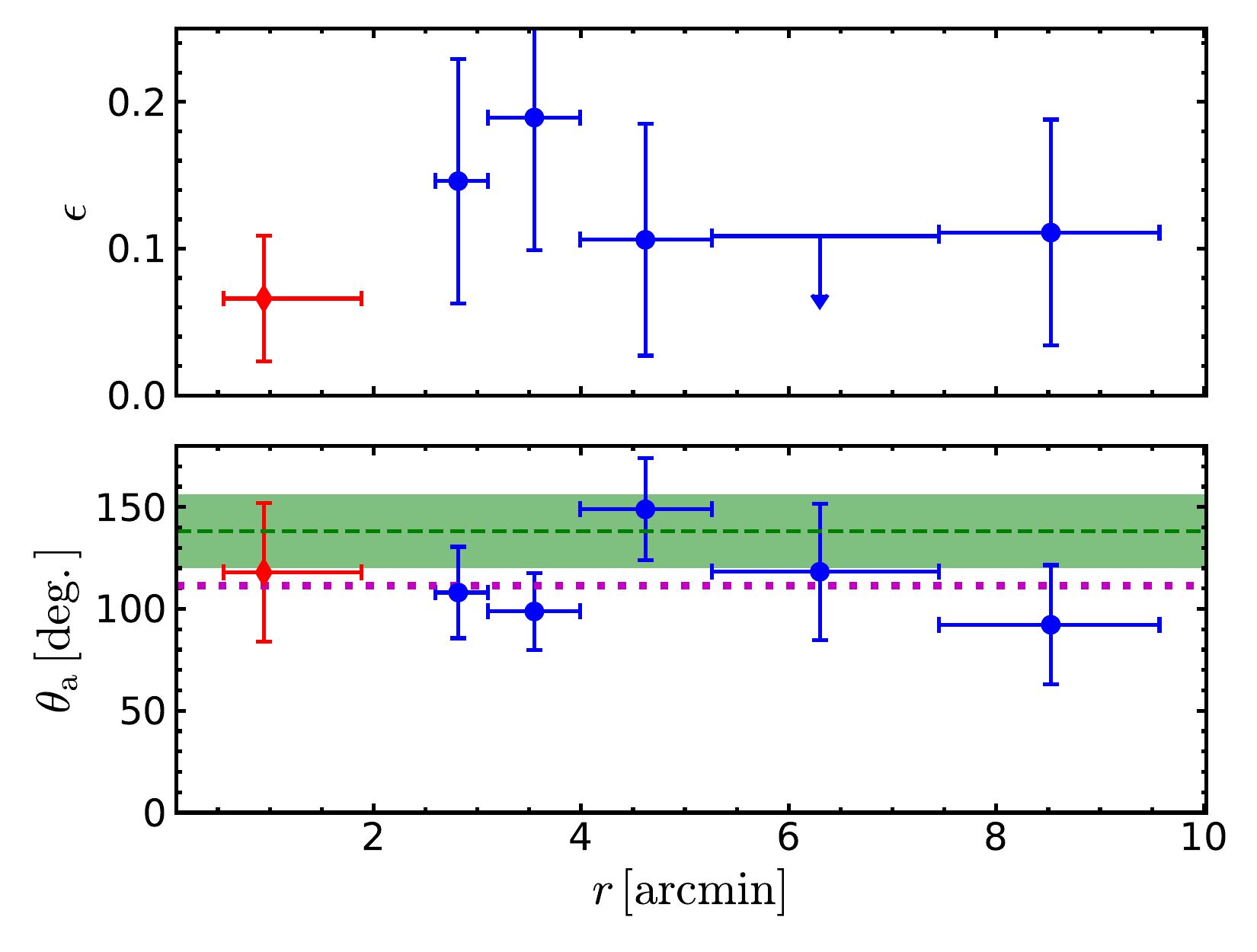}
 \caption{The projected ellipticity ({\it top}) and position angle of the semi-major axis ({\it bottom}, measured from north through east) of NGC~6093 as a function of distance to the cluster centre. In both panels, blue circles indicate results obtained using the catalogue of cluster members from \citet{2019MNRAS.485.4906D} whereas a red diamond indicates the result derived in \citet{2018MNRAS.473.5591K} using HST photometry. In the lower panel, a green dashed line marks the expected position of the semi-major axis for an oblate rotator, assuming the rotation axis angle $\theta_{\rm 0}$ derived in \citet{2018MNRAS.473.5591K}. The direction towards the Galactic Centre is indicated by a dotted magenta line.}
 \label{fig:ellipticity}
\end{figure}

So far, we investigated the morphology of NGC~6093 under the assumption of perfect sphericity. However, there are several effects that can affect the morphology of a cluster, such as rotation, anisotropy, or tidal forces. While the ellipticities of most clusters are still poorly known, {\it Gaia} nowadays provides us with the data necessary to improve the situation. \citet{2019MNRAS.485.4906D} provided source lists of cluster members for all of the objects included in their study. We obtained their source list for NGC~6093 and determined the eigenvalues and eigenvectors of the two-dimensional cluster member distributions in radial bins around the centre \citep[similar to][]{2014ApJ...787L..26F,2018MNRAS.473.5591K}. This allowed us to construct radial profiles of the ellipticity $\epsilon=1-b/a$ and the position angle of the semi-major axis $\theta_{\rm a}$ (measured North through East), which are shown in Fig.~\ref{fig:ellipticity}. Note that we restricted our analysis of the {\it Gaia} data to the region outside the incompleteness limit of $2.59\arcmin$ determined by \citet{2019MNRAS.485.4906D} and used the results from the HST star counts presented in \citet{2018MNRAS.473.5591K} to complement it in the central region. The uncertainties shown in Fig.~\ref{fig:ellipticity} have been calculated by propagating an uncertainty of $2\arcsec$ in the position of the cluster centre and by taking into account the limited number of stars per radial bin \citep[see][for details]{2019MNRAS.483.2197K}.

Our analysis is consistent with a constant ellipticity of $\epsilon\sim0.1$, corresponding to a global axis ratio of $b/a=0.9$. Using 2MASS data, \citet{2010ApJ...721.1790C} determined a global value of $b/a=0.87\pm0.05$, fully consistent with our analysis. The same is true for the position angle, as our measurement of $\theta_{\rm a}=113\pm11^{\circ}$ is fully consistent with the value of $122\pm8^{\circ}$ determined by \citet{2010ApJ...721.1790C}.

In the lower panel of Fig.~\ref{fig:ellipticity}, we also plot the direction in which the cluster would be elongated if it behaved like an oblate rotator. In that case, the semi-major axis should be perpendicular to the rotation axis, which in \citet{2018MNRAS.473.5591K} we found to be at $-132\pm18^{\circ}$ in the central $\sim2\arcmin$. Our analysis is largely consistent with NGC~6093 behaving as an oblate rotator, although there seems to be a slight angular offset between the expected and the true orientation of the semi-major axis. As mentioned earlier, tidal forces can also induce ellipticity. In this case, the cluster is expected to be elongated in direction towards the Galactic Centre. As indicated in Fig.~\ref{fig:ellipticity}, our measurements are consistent with this prediction. However, for NGC~6093, a similar behaviour is predicted from both rotation and tidal forces, so that it is not possible to infer the true origin of the elongation of the cluster. Likely, both effects contribute.

\section{Kinematics}
\label{sec:kinematics}

\begin{figure*}
    \centering
    \includegraphics[width=\textwidth]{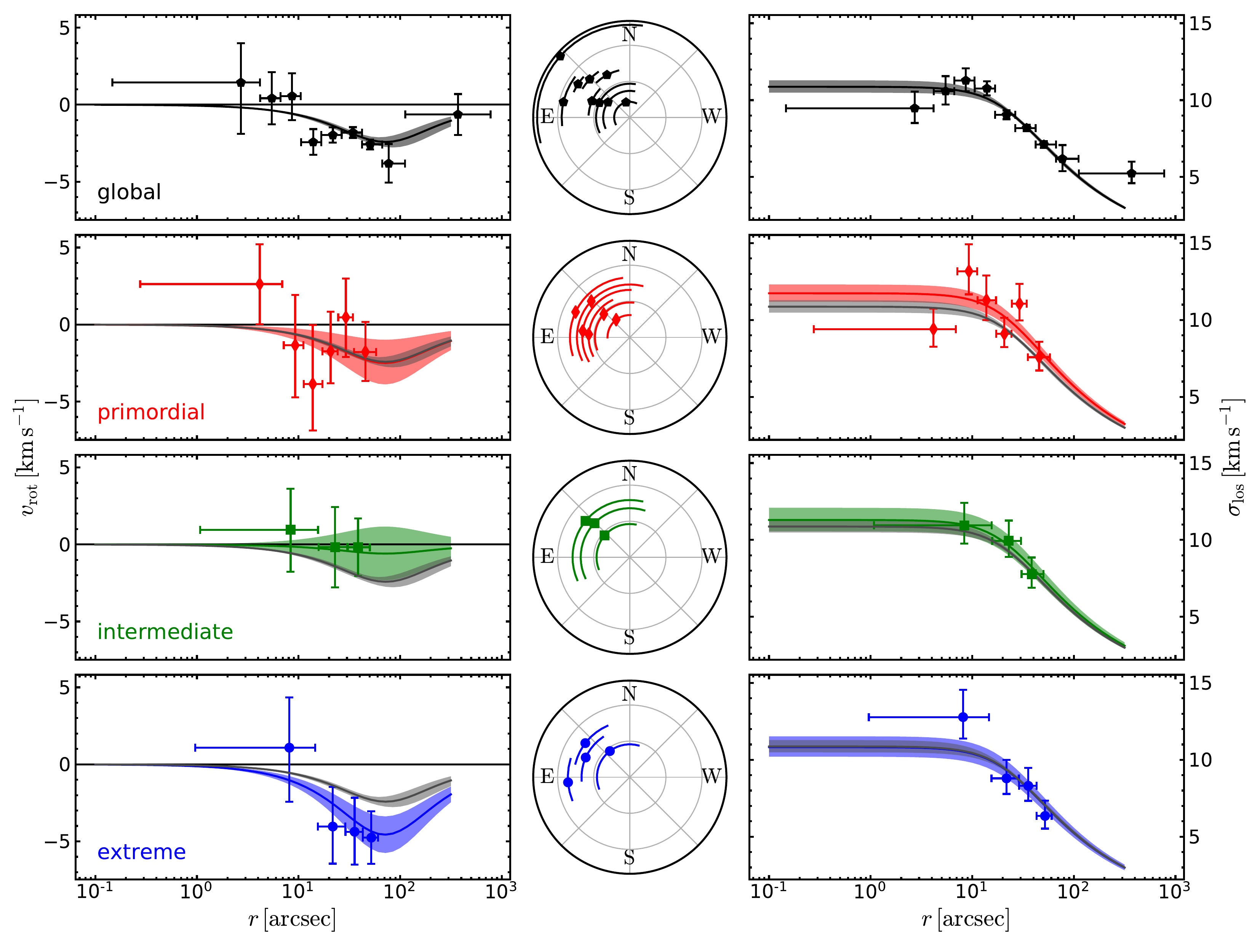}
    \caption{The rotation and dispersion profiles are shown for the entire NGC~6093 kinematic sample and the primordial, intermediate, and extreme populations in the cluster (from top to bottom). Each row shows radial profiles of the rotation amplitude ({\it left}), the orientation of the rotation axis ({\it middle}), and the velocity dispersion ({\it right}). The individual data points correspond to the results obtained in radial bins whereas the continuous profiles were determined via the simple models introduced in Sect.~\ref{sec:kinematics}. Solid lines show the median profiles and shaded areas represent the $1\sigma$ credibility intervals. The model profiles obtained for the full sample are also included in the subsequent rows.}
    \label{fig:population_kinematics}
\end{figure*}

To investigate if the different populations identified by \citet{2018ApJ...859...15D} possess different kinematics, we analysed the velocities of the stars belonging to each population separately and compared them to each other and also to the overall cluster kinematics. As the amount of available MUSE data has increased since our analysis of the same cluster in \citet{2018MNRAS.473.5591K}, we decided to redo the analysis of the overall cluster kinematics. However, we found that the new results are fully consistent to those presented in \citet{2018MNRAS.473.5591K}.

We used the same maximum-likelihood approach as in \citet{2018MNRAS.473.5591K} to infer the rotation velocity $v_{\rm rot}$ and the velocity dispersion $\sigma_{\rm los}$ of the cluster and its individual populations as a function of distance to the cluster centre. It is based on the assumption that at each position ($r$, $\theta$) -- where $r$ is the projected distance towards the cluster centre and $\theta$ is the position angle measured north through east -- the line-of-sight velocity distribution of the cluster can be approximated by a Gaussian with mean value $v_{\rm los} (r, \theta)$ and standard deviation $\sigma_{\rm los}(r)$. In our approach, the variation of the mean with position angle $\theta$ is parametrized according to $v_{\rm los} (r, \theta)=v_{\rm rot}(r)\,\sin(\theta-\theta_{\rm 0})$.

Both the MUSE and the literature samples contain stars not associated with the cluster. To account for these foreground stars, we made use of the cluster membership probabilities $p$ determined in \citet{2018MNRAS.473.5591K} and \citet{2018MNRAS.478.1520B} and modified the likelihood function of each star $i$ according to $\mathcal{L}_i = p_i\mathcal{L}_{{\rm cl}, i} + (1-p_i)\mathcal{L}_{{\rm fg}, i}$, where $\mathcal{L}_{\rm cl}$ and $\mathcal{L}_{\rm fg}$ are the likelihood functions for the cluster and the foreground population, respectively. The likelihood function for the foreground population was constructed as a sum of Gaussian kernels, located at the radial velocities of a mock stellar population generated via the Milky Way model of \citet{2003A&A...409..523R}.

We analysed the kinematics using both a non-parametric and a parametric approach. The former was achieved by binning the data radially, whereas for the latter we employed simple analytical functions to parametrize the radial dependence of $\sigma_{\rm los}$ and $v_{\rm rot}$. More precisely, we adopted a rotation profile of the form
\begin{equation}
    v_{\rm rot}(r) = \frac{2\,v_{\rm max}\,r}{r_{\rm max}}\bigg/\left(1 + \left(\frac{r}{r_{\rm max}}\right)^2\right)\,,
    \label{eq:violent}
\end{equation}
which is characteristic for systems that have undergone violent relaxation \citep{1967MNRAS.136..101L,1973ApJ...186..481G} and is widely used to model the rotation profiles of star clusters \citep[e.g.][]{2018MNRAS.481.2125B}. The dispersion profile was modeled using a \citet{1911MNRAS..71..460P} profile,
\begin{equation}
    \sigma_{\rm los}(r) = \frac{\sigma_{\rm max}}{\left(1 + \left(\frac{r}{a_{\rm 0}}\right)^2\right)^{1/4}}\,.
    \label{eq:plummer}
\end{equation}

In the parametric approach, we determined up to five parameters, $v_{\rm max}$, $r_{\rm max}$, $\theta_{\rm 0}$, $\sigma_{\rm max}$, and $a_{\rm 0}$. When binning the data radially instead, we determined three parameters per bin, $v_{\rm rot}$, $\theta_{\rm 0}$, and $\sigma_{\rm los}$. In both cases, the fitting of the parameters was performed using \textsc{emcee} \citep{2013PASP..125..306F}, which implements the Markov Chain Monte Carlo (MCMC) ensemble sampler from \citet{2010CAMCS...5...65G}. In what follows, we will consider the median values of the distributions sampled by the chains as best-fit parameters and use the 16th and 84th percentiles of said distributions to assign uncertainties. To create radial rotation and dispersion profiles from the parametric models, we randomly drew parameter sets from the chains, evaluated equations~\ref{eq:violent} and \ref{eq:plummer} for each set at a representative set of radii, and obtained the median values as well as the 16th and 84th percentile at each radius.

In one aspect, our analysis deviates from the one performed in \citet{2018MNRAS.473.5591K}, where we used a prior restricting the rotation velocity to positive values and allowed the rotation axis angle $\theta_{\rm 0}$ to take any value in the interval $[-180^{\circ},\,180^{\circ})$. Here, we removed the prior on the rotation velocity and instead restricted $\theta_{\rm 0}$ to the interval $[0^{\circ},\,180^{\circ})$, because each rotation curve with negative axis angle is transitioned into itself when increasing the axis angle by $+180^{\circ}$ and swapping the sign of its amplitude. The advantage of this approach is that for non-rotating clusters, $v_{\rm rot}$ should scatter symmetrically around zero, whereas the previous approach would return a very skewed distribution of rotation velocities. In the case of NGC~6093, where we found an axis angle of $\theta_{\rm 0}=-132\pm18^{\circ}$ in \citet{2018MNRAS.473.5591K}, this implies that the value derived for $\theta_{\rm 0}$ in this work will be shifted by $+180^{\circ}$ on average compared to the previous analysis and the rotation velocities will be $<0$.

In the top panel of Fig.~\ref{fig:population_kinematics}, we show radial profiles of the rotation velocity and the velocity dispersion that we obtained for the full kinematic sample using either the non-parametric or the parametric approach. The best-fit parameters for the parametric approach are presented in the top row of Table~\ref{tab:model_parameters}. We further include in Table~\ref{tab:model_parameters} two measurements of the ordered-over-random motion inside the half-light radius of NGC~6093, namely $(v/\sigma)_{\rm HL}$ and $\lambda_{\rm R,\,HL}$. They were computed from the rotation and dispersion curves depicted in Fig.~\ref{fig:population_kinematics} according to the formulae
\begin{equation}
    \left(\frac{v}{\sigma}\right)_{\rm HL}^2 = \frac{\langle v^2 \rangle}{\langle \sigma_{\rm r}^2 \rangle} = \frac{\int_{0}^{r_{\rm h}} \Sigma(r)\,(1/2)v_{\rm rot}(r)^2\,r\,dr}{\int_{0}^{r_{\rm h}} \Sigma(r)\,\sigma_{\rm los}(r)^2\,r\,dr}
    \label{eq:vsigma}
\end{equation}
and
\begin{equation}
    \lambda_{\rm R,\,HL} = \frac{\langle r |v| \rangle}{\langle r\sqrt{v^2 + \sigma_{\rm r}^2} \rangle} = \frac{\int_0^{r_{\rm h}} \Sigma(r)\,(2/\pi)\lvert v_{\rm rot}(r)\rvert\,r^2\,dr}{\int_0^{r_{\rm h}} \Sigma(r)\sqrt{\sigma_{\rm los}(r)^2+(1/2)v_{\rm rot}(r)^2}\,r^2\,dr}\,.
    \label{eq:lambdar}
\end{equation}
The surface densities $\Sigma(r)$ provided by the MGE models (cf. Table~\ref{tab:mge_parameters}) were used and the half-light radius was adopted as $r_{\rm h}=36\arcsec$ \citep{2018MNRAS.478.1520B}. Note that the factors $(2/\pi)$ and $(1/2)$ under the integrals result from averaging $\lvert v_{\rm los}(r,\,\theta)\rvert$ and $v_{\rm los}(r,\,\theta)^2$ over the position angle $\theta$.

Within the MUSE footprint ($\sim60\arcsec$), our results are in good agreement with the previous analysis, with a central velocity dispersion of $\sim11\,{\rm km\,s^{-1}}$ and a maximum rotation velocity of $\sim3\,{\rm km\,s^{-1}}$. Thanks to the addition of the velocities collected by \citet{2018MNRAS.478.1520B}, we are now able to sample the peak of the rotation curve, which is located at $r_{\rm max}=(1.18^{+0.28}_{-0.23})\arcmin$ (cf. Table~\ref{tab:model_parameters}), corresponding to $2.0\times r_{\rm h}$. The ratio of $r_{\rm max}/r_{\rm h}$ is in agreement with both theoretical models \citep[e.g.][]{2017MNRAS.469..683T} and observations in other clusters \citep[e.g.][]{2018MNRAS.481.2125B}.

In Fig.~\ref{fig:population_kinematics}, we also show the rotation and dispersion profiles derived for the primordial, intermediate, and extreme population in NGC~6093. As for the overall sample, we provide the best-fit parameters of the parametric approach in Table~\ref{tab:model_parameters}. Note that because of the limited radial coverage and the smaller sample sizes, we did not try to constrain the radial scales ($r_{\rm peak}$ and $a_{\rm 0}$) of the rotation and dispersion curves, but fixed them to the values obtained for the full sample instead.

\begin{table*}
	\centering
	\caption{Best-fit parameters of the rotation and dispersion models (cf. equations~\ref{eq:violent},~\ref{eq:plummer}) for the overall sample and the different stellar populations in NGC~6093.}
	\label{tab:model_parameters}
    \include{ngc6093_population_kinematics}
\end{table*}

We obtain velocity dispersion profiles that are in agreement with the one derived for the whole cluster. However, at radii $\gtrsim10\arcsec$, there is a slight trend that the dispersion decreases when going from the primordial to the intermediate to the extreme population. While the significance of this trend is low, it agrees with the expectations. As the gravitational potential is the same for all three populations, the dispersion should scale with the observed concentration, which is highest (lowest) for the primordial (extreme) population. On the other hand, inside $10\arcsec$, the results are ambiguous. While the results from the parametric approach are consistent with those at larger radii, we we find a higher dispersion for the central bin of the extreme population than for the central bin of the primordial population. However, as explained in Sect.~\ref{sec:analysis}, our capabilities to measure radial velocities for stars in either population are affected by the extreme crowding in this radial range. For this reason, the central dispersion measurements are the most uncertain. We will revisit the velocity dispersion profiles in Sect.~\ref{sec:models} below.

The rotation profiles show a remarkable behaviour across the populations. While no differences are observed regarding the orientations of the rotation axes, which for all three populations are consistent with the global one, the opposite is true for the strengths of the rotation fields. The extreme population shows the strongest rotation, with a peak amplitude of $\lvert v_{\rm max}\rvert = 4.52\pm1.17\,{\rm km\,s^{-1}}$.  On the other hand, almost no rotation is observed for the stars of the intermediate population, for which our analysis yields an upper limit of $\lvert v_{\rm max}\rvert < 2.28\,{\rm km\,s^{-1}}$. The primordial population seems to lie in between the two other populations in terms of rotation amplitude, with a value of $\lvert v_{\rm max}\rvert=2.49^{+1.34}_{-1.45}\,{\rm km\,s^{-1}}$. It should be noted that due to the limited number of available stars per population, the uncertainties of the derived parameters are relatively large compared to those obtained for the entire cluster. Nevertheless, we find the difference in rotation velocity between the intermediate and extreme population to be significant at the $>2\sigma$ level, with $96.7\%$ of the MCMC samples yielding a higher rotation velocity for the extreme population. On the other hand, the differences between the primordial population and either the intermediate or the extreme population are only significant at the $>1\sigma$ level, with $79.4\%$ of the samples favouring a higher rotation velocity in the primordial than in the intermediate population and $87.1\%$ of the samples favouring a higher rotation velocity in the extreme than in the primordial population. We note that our results are in qualitative agreement with the study of M13 by \citet{2017MNRAS.465.3515C}, who also found the highest rotation velocity for the extreme population in that cluster.

When comparing the values of $(v/\sigma)_{\rm HL}$ and $\lambda_{\rm R,\,HL}$ of the different populations (cf. Table~\ref{tab:model_parameters}), we see that the degree of ordered-over-random motion in the primordial population is comparable to the global one. On the other hand, lower and higher degrees are found for the intermediate and the extreme populations, respectively. This is in agreement with the conclusions drawn from the rotation curves alone. The differences in $(v/\sigma)_{\rm HL}$ and $\lambda_{\rm R,\,HL}$ are slightly more significant than those in $v_{\rm max}$. The reason is that equations~\ref{eq:vsigma} and \ref{eq:lambdar} also take the different concentrations of the populations into account.

Finally, we investigated whether forcing the rotation profiles of all three populations to the same scale radius $r_{\rm max}$ found for the global sample has a significant impact on our results. To this aim, we repeated the analysis for the three populations and allowed $r_{\rm max}$ to vary. We found the scale radii to be essentially unconstrained, with the possible exception of the extreme population, where the a posteriori distribution of $r_{\rm max}$ values showed a peak at around $0.7\arcmin$ overlaid on a uniform distribution. The reason for this behaviour is that our population data only cover the rising flank of the profile described by eq.~\ref{eq:violent}, hence shifting the scale radius towards larger values can be compensated by assuming a larger peak velocity $v_{\rm max}$. Consequently, the a posteriori distributions of $v_{\rm max}$ become skewed towards large (absolute) values, and the differences between the populations reported above are partially washed out. However, we also find that both $(v/\sigma)_{\rm HL}$ and $\lambda_{\rm R,\,HL}$ are robust against the change in our analysis setup, as we obtain values in very good agreement with those listed in Table~\ref{tab:model_parameters}. For example, the updated values of $\lambda_{\rm R,\,HL}$ are $0.051^{+0.030}_{-0.030}$, $0.024^{+0.028}_{-0.018}$, and $0.122^{+0.057}_{-0.035}$ for the primordial, intermediate, and extreme populations, respectively. This highlights that $(v/\sigma)_{\rm HL}$ and $\lambda_{\rm R,\,HL}$ can be considerably more useful to characterize the kinematics of a given population than individual parameters of physically motivated analytical profiles like those given in equations~\ref{eq:violent} and \ref{eq:plummer}.

\section{Dynamical models}
\label{sec:models}

As the different populations identified in NGC~6093 have different radial distributions, their observed kinematics are expected to be different as well. Even when sampled at the same projected radii, the intrinsic distributions of the stars along the line of sight will vary depending on the population under review. To quantify this effect and to verify if it can be the sole explanation for the differences observed in Sect.~\ref{sec:kinematics}, we use axisymmetric Jeans models. Such models predict the first- and second-order velocity moments as a function of position for a stellar system that can be both rotating and anisotropic. The formalism behind the models as well as the underlying assumptions are explained in \citet{2008MNRAS.390...71C}. The \textsc{cjam} software used in this work was presented in \citet{2013MNRAS.436.2598W}. It uses the following properties as input:
\begin{itemize}
    \item To predict the velocity moments, the code requires the projected number density of the tracer population (i.e. the population for which the kinematics have been measured) in the form of an elliptical MGE. We used the parametrizations listed in Table~\ref{tab:mge_parameters} for this purpose, and accounted for the projected ellipticity of NGC~6093 (cf. Fig.~\ref{fig:ellipticity}) by assigning each Gaussian component an axis ratio of $q=0.9$. The $\sigma_{k}$ values listed in Table~\ref{tab:mge_parameters} were divided by $q$ to obtain the standard deviations along the semi-major axes of the components.
    \item The gravitational potential is derived from a projected mass density profile, which must also be provided in the form of an elliptical MGE. We used a scaled version of the parametrization of the global profile given in Table~\ref{tab:mge_parameters}. As described in Sect.~\ref{sec:morphology}, we converted the projected number density profile to a projected luminosity density profile by requiring that the apparent cluster magnitude from \citet{1996AJ....112.1487H} is recovered when integrating over it. Then, the projected mass density profile is obtained by applying a scaling factor $\Upsilon$ -- equivalent to a mass-to-light ratio -- to each Gaussian component. Again, we adopted a constant axis ratio of $q=0.9$ for all components.
    \item As distance to the cluster, we assumed $d=10\,{\rm kpc}$ \citep{1996AJ....112.1487H}. Note that in the absence of any proper motions, the distance $d$ and the total mass-to-light ratio $\Upsilon$ cannot be constrained at the same time. For this reason, we did not try to constrain $d$ via the models.
    \item The Jeans models adapted here implement rotation via a parameter $\kappa$, which is zero in the absence of rotation and otherwise scales with the strength of the rotation field. A separate value of $\kappa$ can be assigned to each Gaussian component of the \emph{tracer} population.
    \item The global inclination of the system can vary from $i=0^{\circ}$ for face-on systems to $i=90^{\circ}$ for edge-on systems. Following \citet{2013MNRAS.436.2598W}, we optimized our models for the median intrinsic flattening ${\overline q}$ of the MGE, and calculated $i$ afterwards via $\cos i = \sqrt{\frac{q^2 - {\overline q}^2}{1 - {\overline q}^2}}$.
    \item Finally, the models handle anisotropy via a parameter $\beta$. However, constraining anisotropies from radial velocities alone is very challenging. Therefore, we assumed isotropy ($\beta\equiv0$) in all our analyses. This is a reasonable assumption at least for the central part of NGC~6093, given that the half-light relaxation time of NGC~6093 \citep[$\log (t_{\rm h}/{\rm yr}) = 8.8$,][]{1996AJ....112.1487H} is in a regime where other clusters appear almost isotropic \citep{2015ApJ...803...29W}.
\end{itemize}

Provided a given set of model parameters, the code yields the first and second order velocity moments for a user-defined set of spatial positions ($x^\prime$, $y^\prime$), with the $x^\prime$-axis being aligned with the projected semi-major axis of the system. The moments are returned for three dimensions, which are oriented along the line of sight (i.e. the $z^\prime$-axis) and in the $x^\prime$- and $y^\prime$-directions. In the absence of proper motions, we only used the components along the line of sight, $\overline{v_{z^\prime}}$ and $\overline{v^2_{z^\prime}}$, however.

To compare our kinematic data to the models, we applied a rotation by an angle of $58\,{\rm deg.}$ (cf. Table~\ref{tab:model_parameters}) to the data, so as to align the rotation axis (i.e. the semi-minor axis) with the model $y^\prime$-axis. By using the moments $\overline{v_{z^\prime}}$ and $\overline{v^2_{z^\prime}}$ returned by the code, we were then able to determine a likelihood for each model given the data. The formulae for calculating the likelihood can be found in \citet{2013MNRAS.436.2598W} and are not repeated here. As in Sect.~\ref{sec:kinematics}, we used \textsc{emcee} to find the best-fitting parameters for each of our data sets.

\subsection{Global cluster model}

\begin{table*}
	\centering
	\caption{Best-fit parameters of the Jeans models used to describe the overall sample and the different stellar populations in NGC~6093.}
	\label{tab:cjam_parameters}
    \include{ngc6093_population_models}
\end{table*}

To fit the full set of radial velocities in NGC~6093, we optimized the models in terms of the scaling factors $\Upsilon_n$, the rotation parameters $\kappa_n$, and the inclination angle $i$. As $6$ MGE components were required to fit the profile of NGC~6093 (cf. Sect.~\ref{sec:morphology}), this leaves us with $6\times2+1=13$ model parameters to optimize. However, given the overlap of the Gaussian components in radius, we expect some degeneracies between the individual values of $\Upsilon_n$ and $\kappa_n$. Therefore, we decided to parametrize $\Upsilon(r)$ as 
\begin{equation}
    \Upsilon(r) = \frac{\Upsilon_0 \left[ 1 - \left(\frac{r}{r_{\rm \Upsilon}}\right)\right] + 2 \Upsilon_{\rm t}\left(\frac{r}{r_{\rm \Upsilon}}\right) + \Upsilon_{\infty} \left(\frac{r}{r_{\rm \Upsilon}}\right) \left[1 - \left(\frac{r}{r_{\rm \Upsilon}}\right)\right]}{1+\left(\frac{r}{r_{\rm \Upsilon}}\right)^2}\,,
    \label{eq:upsilon}
\end{equation}
and $\kappa(r)$ as
\begin{equation}
    \kappa(r) =  \frac{2 \kappa_{\rm max}\left(\frac{r}{r_{\rm \kappa}}\right)}{1 + \left(\frac{r}{r_{\rm \kappa}}\right)^2}\,.
    \label{eq:kappa}
\end{equation}
While eq.~\ref{eq:kappa} is basically an adoption of the violent relaxation model (eq.~\ref{eq:violent}) for $\kappa$, eq.~\ref{eq:upsilon} describes a function which approaches $\Upsilon_0$ for $r=0$, $\Upsilon_\infty$ for $r\rightarrow\infty$, and takes a transition value $\Upsilon_t$ at $r=r_{\rm \Upsilon}$. This choice of function is motivated by the observation that many globular clusters possess a well-defined minimum in their mass-to-light ratio profiles \citep{2017MNRAS.464.2174B}. To assign values to the individual MGE components, we evaluated eqs~\ref{eq:upsilon} and \ref{eq:kappa} at the radii where the components' contributions to the global profile were maximal. An alternative approach would be to optimize only a subset of the $\Upsilon_m$ and $\kappa_n$ and obtain the remaining values via interpolation \citep[e.g.][]{2014MNRAS.438..487D}. We verified that the final results were insensitive to the approach adopted.

The best-fit parameters we obtained from the maximum-likelihood analysis are listed in Table~\ref{tab:cjam_parameters}. The lower limit on the intrinsic flattening of ${\overline q} = 0.78$ corresponds to a minimum inclination angle of $i\geq\,45{\rm deg}$, hence the cluster is seen preferentially edge-on. This could provide an explanation to the non-detection of rotation in NGC~6093 in the recent proper motion study of \citet{2018MNRAS.481.2125B}.

\begin{figure}
 \includegraphics[width=\columnwidth]{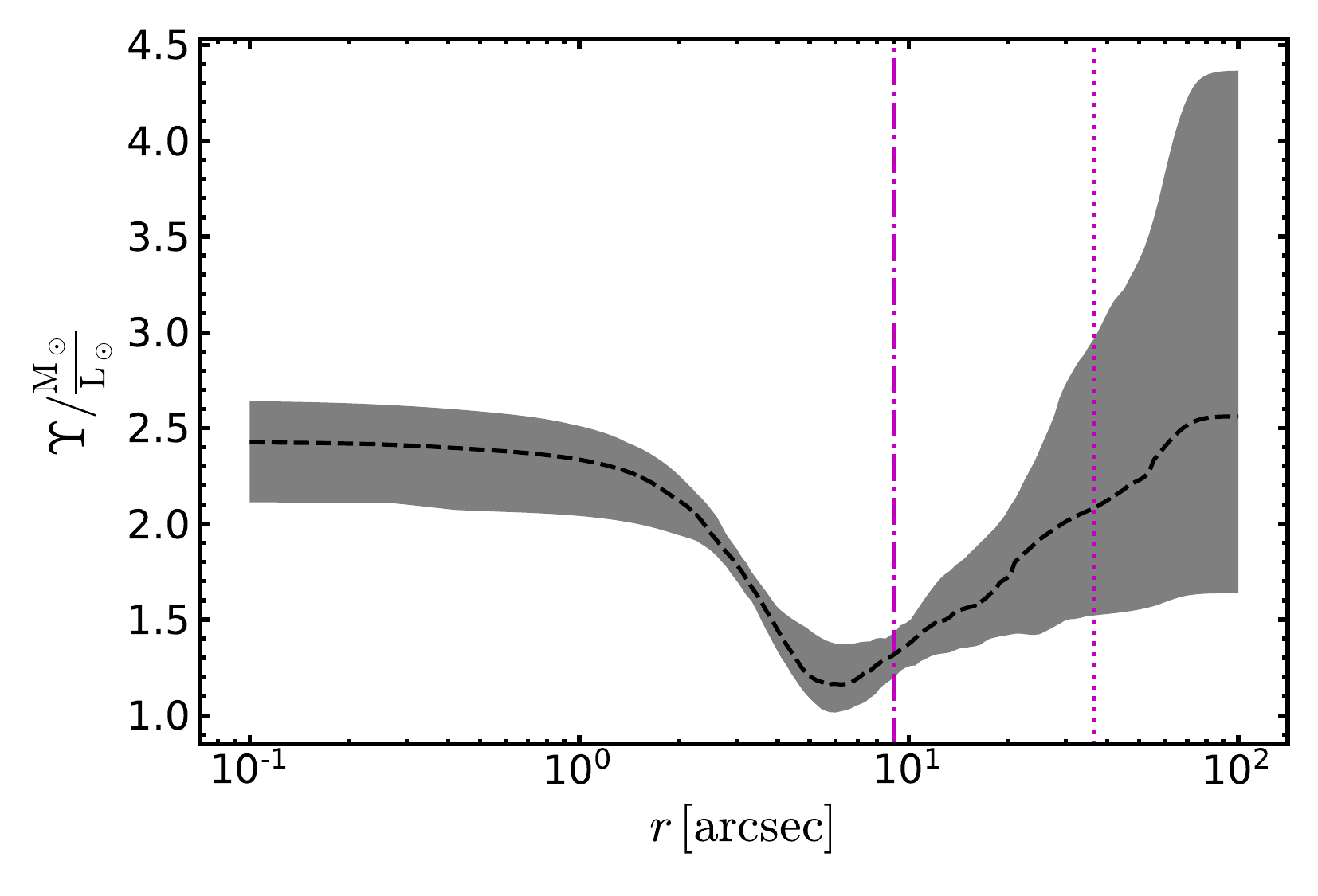}
 \caption{The projected mass-to-light ratio of NGC~6093 as a function of radius. The dashed line indicates the median of 100 profiles drawn randomly from the MCMC chain. The grey-shaded area encompasses 68\% of said profiles. Vertical dotted and dash-dotted line indicate the half-light and the core radius of the cluster, respectively.}
 \label{fig:mlr_profile} 
\end{figure}

The parameters going into eq.~\ref{eq:upsilon} listed in the top row of Table~\ref{tab:cjam_parameters} suggest that the mass-to-light ratio of NGC~6093 does have a minimum at around $0.2\arcmin$. To verify this, we randomly drew parameter sets from the MCMC chain and determined the mass-to-light ratio profile corresponding to each of them. In Fig.~\ref{fig:mlr_profile}, we show the median profile and the associated uncertainty interval, confirming the minimum. The shape is in qualitative agreement with the mean mass-to-light ratio of the clusters investigated by \citet{2017MNRAS.464.2174B}, which displayed a minimum value of $\sim1.2\,{\rm M_\odot\,L_\odot^{-1}}$ at around $(0.1-0.2)\times$ the half-light radius. We obtain a global mass-to-light ratio of $\Upsilon_{\rm c}=1.72\pm0.20\,{\rm M_\odot\,L_\odot^{-1}}$, corresponding to a cluster mass of $M_{\rm c}=2.85\pm0.34\times10^5\,{\rm M_\odot}$. This cluster mass is in good agreement with the results of the latest N-body models from \citet{2018MNRAS.478.1520B}, yielding $(2.79\pm0.08)\times10^5\,{\rm M_\odot}.$\footnote{see \url{https://people.smp.uq.edu.au/HolgerBaumgardt/globular/fits/ngc6093.html}.}

\subsection{Individual populations}

\begin{figure}
    \centering
    \includegraphics[width=\columnwidth]{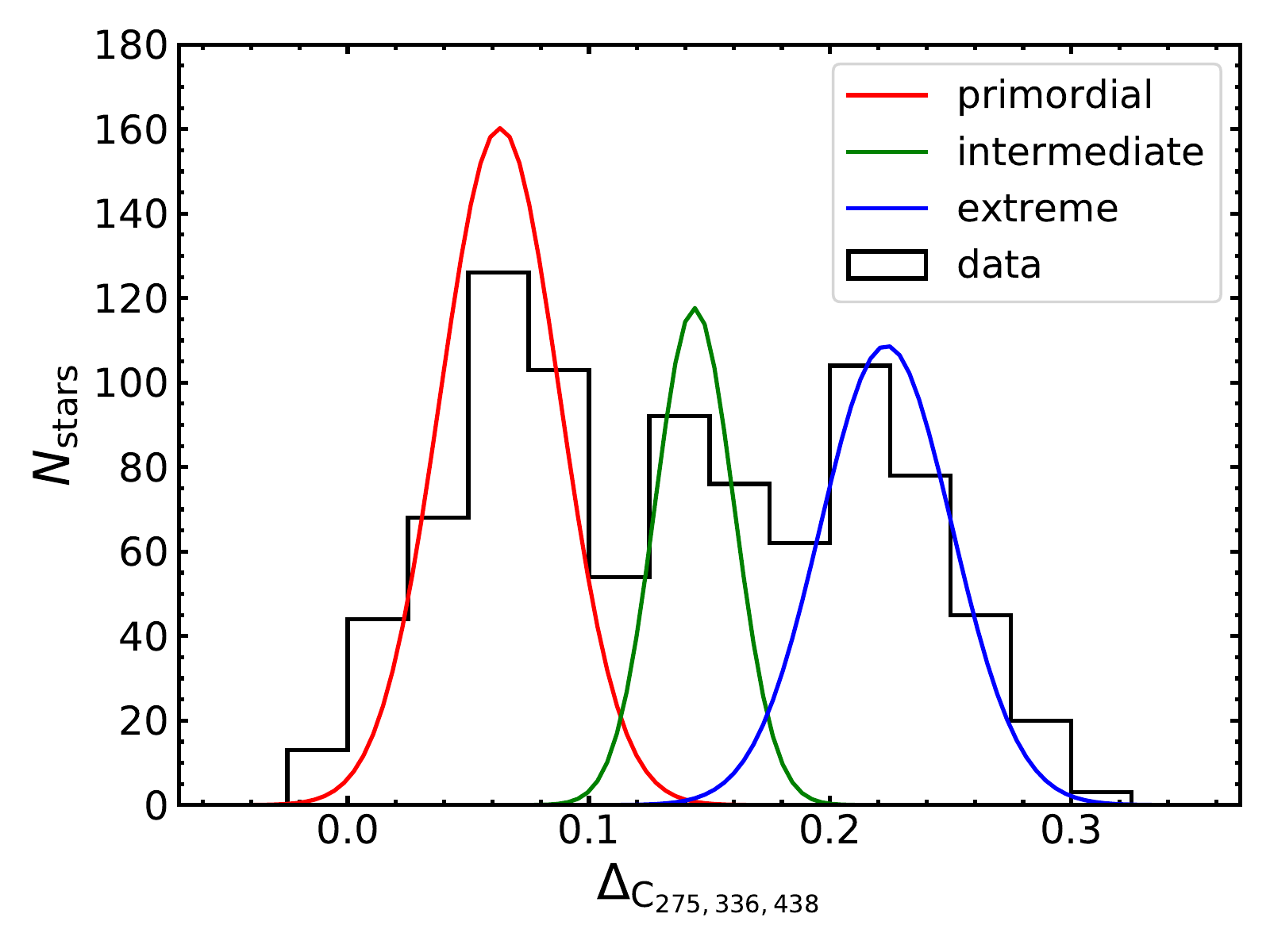}
    \caption{Comparison between the distribution of pseudo-colours along the red giant branch of NGC~6093 and the results of the multi-Gaussian model adopted in the Jeans models. Note that in contrast to the data, the Gaussian components are not broadened by the measurement uncertainties.}
    \label{fig:pseudo_colour_fit}
\end{figure}

\begin{figure*}
    \centering
    \includegraphics[width=\textwidth]{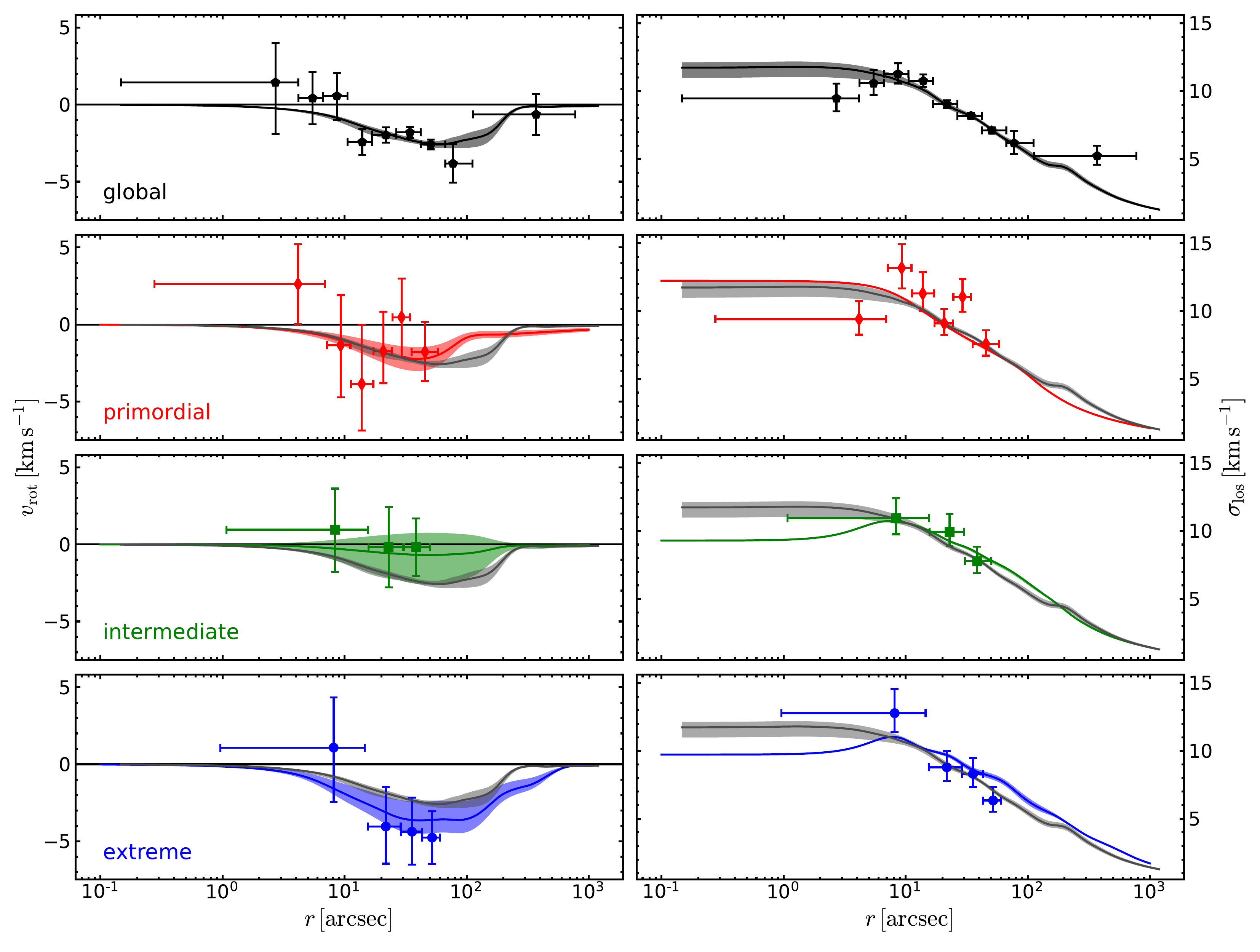}
    \caption{Comparison between the binned rotation and dispersion profiles and the predictions from axisymmetric Jeans models for the entire NGC~6093 sample and the primordial, intermediate, and extreme populations in the cluster (from top to bottom). Each row shows radial profiles of the rotation amplitude ({\it left}) and the velocity dispersion ({\it right}). The model predictions obtained for the full sample are also included in the subsequent rows as grey-shaded areas.}
    \label{fig:population_kinematics_cjam}
\end{figure*}

With a model for the entire cluster at hand, we are able to investigate the behaviour of its various subpopulations. As was already mentioned, the gravitational potential remains the same, irrespective of which population is studied. For this reason we choose to leave the MGE describing the projected mass density of NGC~6093 unchanged and fixed the parameters affecting its scaling factors, $\Upsilon_{\rm 0}$, $\Upsilon_{\rm t}$, $\Upsilon_{\infty}$, and $r_{\rm \Upsilon}$, to their median values obtained in the previous MCMC analysis (cf. Table~\ref{tab:cjam_parameters}). In addition, the intrinsic flattening $\overline{q}$ was fixed to the value obtained for the global model. Finally, in light of the limited radial coverage of the stars for which a population tag is available, we fixed $r_\kappa$ for each population to the median value found for the entire cluster (cf. Table~\ref{tab:cjam_parameters}). Hence, the only dynamical parameter to be determined for each population is $\kappa_{\rm max}$, which we consider as a proxy for the angular momentum of each population.

A straightforward approach of inferring $\kappa_{\rm max}$ for each population would be to determine the best-fit Jeans model for each of the kinematic subsamples used in Sect.~\ref{sec:kinematics} independently. However, a downside of this approach is that we would neglect the information provided by the stars without a population tag, i.e. those 174 out of 733 stars shown as grey dots in the inset panel in Fig.~\ref{fig:cmd}. In order to include those stars in the modelling, we decided to create a single chemo-dynamical model that accounts for all three populations simultaneously. To this aim, besides the MUSE velocity measurements, we also used the $\Delta_{\rm C_{275,336,438}}$ pseudo-colours (and the associated uncertainties) determined by \citet[see their Fig.~3]{2018ApJ...859...15D} and assumed that the intrinsic distribution of pseudo-colours in each population could be approximated as a Gaussian with parameters $\mu_{\rm \Delta_C}$ and $\sigma_{\rm \Delta_C}$. We then added three components to the model, each of which was characterized by three parameters, $\kappa_{\rm max}$, $\mu_{\rm \Delta_C}$, and $\sigma_{\rm \Delta_C}$. Hence in total nine parameters were to be optimized in a single MCMC run.

A similar approach has been pursued by \citet{2016MNRAS.463.1117Z} in their analysis of the Sculptor dwarf galaxy, but using two instead of three subpopulations. \citet{2016MNRAS.463.1117Z} also used their Jeans model to constrain the spatial density profiles of the populations, by using linear interpolations of the MGE profiles determined photometrically and constraining the interpolation coefficients during the MCMC run. While we looked into the feasibility of adapting this idea in our analysis of NGC~6093, we found that in our case the interpolation coefficients were very poorly constrained, presumably because the populations show less pronounced density profile differences than is the case for Sculptor. Therefore, we adopted the \emph {tracer} MGEs found in Sect.~\ref{sec:morphology} to describe the spatial distributions of the three populations.

The results from the Jeans modeling of the populations are illustrated in Figs~\ref{fig:pseudo_colour_fit} and \ref{fig:population_kinematics_cjam}, while the best-fit parameters are summarized in Table~\ref{tab:cjam_parameters}. As can be verified from Fig.~\ref{fig:pseudo_colour_fit}, where we compare the Gaussian $\Delta_{\rm C_{275,336,438}}$ pseudo-colour distributions obtained from our model to the actual data, we can accurately describe the complex chemistry of NGC~6093. Note that in contrast to the data, the Gaussian curves are not broadened by measurement uncertainties, but only account for the intrinsic width of each population in pseudo-colour space. Hence, they appear somewhat narrower than the data. The fact that for each of the populations we find a non-negligible intrinsic spread (see also the $\sigma_{\rm \Delta_C}$ values listed in Table~\ref{tab:cjam_parameters}) indicates that none of the populations is actually completely homogeneous in its chemistry, but that slight abundance variations also exist within each population.

Turning our focus to the kinematics of the populations, we find that the confidence intervals of the radial velocity dispersion profiles are very narrow, see the right panels of Fig.~\ref{fig:population_kinematics_cjam}. The reason for this is that the observed velocity second order moment $\overline{v^2}=\sqrt{{\overline v}^2 + \sigma^2}$ is essentially defined by the gravitational potential and the projected distribution of the tracer population, both of which are not varied in our models. Hence, only the contributions of ordered and random motions to $\overline{v^2}$ change, and even for a rotating cluster such as NGC~6093, ${\overline v}^2 \ll \sigma^2$ typically, so that $\sigma$ is hardly affected by changes to ${\overline v}$.

We also see in Fig.~\ref{fig:population_kinematics_cjam} that the different radial distributions of the populations alone have only a minor impact on the observed kinematics, at least in the radial range that we can probe with the MUSE data. The Jeans models predict differences between the dispersion profiles of the individual populations and the global one of typically $<1\,{\rm km\,s^{-1}}$. This is somewhat smaller than our measurements, as can be seen by the slight underestimation of the dispersion profile measured for the primordial population. A possible explanation for this discrepancy could be that some of our model assumptions (e.g. isotropy) are violated in the kinematics of said population. In the inner $\sim10\arcsec$, the Jeans models predict more significant differences between the populations, in particular a dip in the dispersion profiles of the intermediate and extreme population. Unfortunately, our recovery of stars from the catalog of \citet{2018ApJ...859...15D} is highly incomplete in this region, so we cannot verify the existence of such dips.

The Jeans models are also able to provide accurate representations of the rotation profiles of the individual populations, as can be verified in the left panels of Fig.~\ref{fig:population_kinematics_cjam}. Furthermore, the best-fit values listed in Table~\ref{tab:cjam_parameters} show that the models converge to different $\kappa_{\rm max}$ values. The modeling results strengthen the finding of Sect.~\ref{sec:kinematics} that the extreme population possesses a higher degree of ordered motions than the intermediate one. In $95.0\%$ of the MCMC samples, the $\kappa$-value of the extreme population is larger than that of the intermediate population. In addition, we find with similar significance ($93.1\%$) that the extreme population also rotates faster than the primordial population. On the other hand, the differences between the rotation of the primordial and intermediate populations are less significant, with $76.4\%$ of the MCMC samples returning a higher $\kappa$ for the primordial population. The differences in ordered-to-random motion between the populations that are suggested by the models can also be verified by the $\lambda_{\rm R,\,HL}$ values included in Table~\ref{tab:cjam_parameters}.

\section{Conclusions}
\label{sec:conclusions}

We presented a study of the chemistry and dynamics of the globular cluster NGC~6093 by combining MUSE integral field spectroscopy with \textit{HST} photometry. For each of the three populations identified in the photometry, we were able to derive the radial velocities of a sufficiently large sample of stars for a kinematic analysis. By using axisymmetric Jeans models as well as simple parametric and non-parametric radial profiles, we found that all three populations are rotating, with consistent (projected) rotation axis orientations. However, the the stellar population with the highest N-enrichment (i.e. the extreme population) has a higher projected rotation velocity than the remaining two populations. The comparison to the axisymmetric Jeans models showed that the observations can be explained under the assumption that the populations differ in their intrinsic rotation properties and have different angular momenta. No process is known that could induce such variations during the evolution of the cluster, hence we assume that they were imprinted during the formation of NGC~6093.

While a higher rotation of the N-enriched population matches the predictions for formation scenarios based on multiple epochs of star formation, it is difficult to reconcile with the suggested advanced dynamical stage of the cluster \citep{2012Natur.492..393F}. If the central concentrations of the populations have indeed been flipped by mass segregation, as suggested by \citet{2018ApJ...859...15D}, it seems very unlikely that the kinematic differences imprinted at the birth of the cluster have survived such a high degree of relaxation. On the other hand, mass segregation imposes changes in the second velocity moment $\langle v^2\rangle$, not necessarily in the first moment $\langle v\rangle$. Therefore, dedicated simulations will need to be performed to investigate this further. While the N-body models used by \citet{2018ApJ...859...15D} to investigate differences in the populations' radial distributions did not include rotation, we plan to investigate this aspect further with dedicated simulations.

Our findings are remarkably similar to those of \citet{2017MNRAS.465.3515C}, who found the extreme population of the cluster NGC~6205 to be rotating faster than the remaining populations. Similar to NGC~6093, an advanced dynamical state has also been suggested for NGC~6205 \citep[see][]{2018MNRAS.474.4438S}. Given that the Milky Way hosts further clusters that harbour populations with abundance patterns comparable to NGC~6093 and NGC~6205, these clusters appear as promising candidates to confirm or refute if enhanced rotation is a general property of the extreme population.

We sound a note of caution in that due to the relatively low number of stars with population tags, the significance of the observed differences is still limited. One opportunity to improve on this would be to increase the number of stars with kinematic data at larger radii. The available MUSE data cover only part of the footprint of the \textit{HST} photometry used by \citet{2018ApJ...859...15D} and $\sim 4$ additional pointings would be a possibility to complete the coverage. However, as visible from Fig.~\ref{fig:surface_brightness}, the HST coverage stops inside of $100\arcsec$, where the largest differences are expected (cf. Fig.~\ref{fig:population_kinematics_cjam}). High resolution spectra would provide an obvious opportunity to determine the kinematics and the chemistry of stars at radii where no \textit{HST} data are available. Based on the current sample sizes, we estimate that $\gtrsim500$ additional radial velocities would be required to determine the significance of the measured differences at $>3\sigma$ confidence. Alternatively, combining ground-based photometry with \emph{Gaia} data could offer a possibility to extend the analysis to such radii \citep[see][]{2018MNRAS.479.5005M}. However, as our analysis suggests that the rotation field of NGC~6093 is seen preferentially edge-on, the signal in the proper motions may be small \citep[as suggested by the lack of rotation in the data of][]{2018MNRAS.481.2125B}.


\section*{Acknowledgements}
We thank the anonymous referee for their helpful report.
Based on observations made with ESO Telescopes at the La Silla Paranal Observatory under programme IDs 095.D-0629(A), 098.D-0148(A), and 099.D-0019(A).
SK and NB gratefully acknowledge funding from a European Research Council consolidator grant (ERC-CoG-646928- Multi-Pop).
ED acknowledges support from The Leverhulme Trust Visiting Professorship Programme VP2-2017-030.
NB acknowledges support from the Royal Society in the form of a University Research Fellowship.
SD, BG, FG, and TOH acknowledge funding from the Deutsche Forschungsgemeinschaft (grant DR 281/35-1 and KA 4537/2-1) and from the German Ministry for Education and Science (BMBF Verbundforschung) through grants 05A14MGA, 05A17MGA, 05A14BAC, and 05A17BAA.
GvdV and LLW acknowledge funding from the European Research Council (ERC) under the European Union's Horizon 2020 research and innovation programme under grant agreement No 724857 (Consolidator Grant ArcheoDyn).
JB acknowledges support by FCT/MCTES through national funds (PIDDAC) by grant UID/FIS/04434/2019 and through Investigador FCT Contract No. IF/01654/2014/CP1215/CT0003.




\bibliographystyle{mnras}
\bibliography{m80_chemo_dynamics} 







\bsp	
\label{lastpage}
\end{document}

%% file: ngc6093_mge_parameters.tex
\begin{tabular}{l cc cc cc cc}
\hline
 & \multicolumn{2}{c}{global} & \multicolumn{2}{c}{primordial} & \multicolumn{2}{c}{intermediate} & \multicolumn{2}{c}{extreme}\\
$k$ & $\sigma_{k}$ & $\log\Sigma_{k, {\rm 0}}$ & $\sigma_{k}$ & $\log\Sigma_{k, {\rm 0}}$ & $\sigma_{k}$ & $\log\Sigma_{k, {\rm 0}}$ & $\sigma_{k}$ & $\log\Sigma_{k, {\rm 0}}$\\
 & ${\rm arcsec}$ & ${\rm arcsec}^{-2}$ & ${\rm arcsec}$ & ${\rm arcsec}^{-2}$ & ${\rm arcsec}$ & ${\rm arcsec}^{-2}$ & ${\rm arcsec}$ & ${\rm arcsec}^{-2}$\\
\hline
$1$ & $0.7308$ & $3.059$ & $3.089$ & $-1.724$ & $2.556$ & $-0.743$ & $3.165$ & $-0.647$\\
$2$ & $5.697$ & $4.513$ & $5.147$ & $-0.863$ & $5.293$ & $-0.866$ & $7.849$ & $-1.003$\\
$3$ & $13.56$ & $4.161$ & $8.613$ & $-0.873$ & $11.53$ & $-1.400$ & $21.89$ & $-1.674$\\
$4$ & $31.45$ & $3.544$ & $14.61$ & $-1.206$ & $25.18$ & $-1.970$ & $61.98$ & $-2.257$\\
$5$ & $60.53$ & $2.957$ & $25.39$ & $-1.649$ & $51.21$ & $-2.524$ & $151.1$ & $-2.810$\\
$6$ & $158.3$ & $1.453$ & $45.62$ & $-2.258$ & $96.3$ & $-3.142$ & $301.8$ & $-3.475$\\
\hline
\end{tabular}

%% file: ngc6093_population_kinematics.tex
\begin{tabular}{cccccccc}
\hline
population & $\sigma_{\rm max}$ & $v_{\rm max}$ & $\theta_{\rm 0}$ & $r_{\rm max}$ & $a_{\rm 0}$ & $(v/\sigma)_{\rm HL}$ & $\lambda_{\rm R,\,HL}$ \\
& $\mathrm{km\,s^{-1}}$ & $\mathrm{km\,s^{-1}}$ & $\mathrm{deg}$ & $\mathrm{arcmin}$ & $\mathrm{arcmin}$ &  & \\
\hline
 global & $10.87^{+0.42}_{-0.38}$ & $-2.45^{+0.30}_{-0.35}$ & $58^{+3}_{-3}$ & $1.18^{+0.28}_{-0.23}$ & $0.40^{+0.05}_{-0.05}$ & $0.079^{+0.010}_{-0.008}$ & $0.064^{+0.009}_{-0.007}$ \\[5pt]
 primordial & $11.74^{+0.57}_{-0.55}$ & $-2.50^{+1.51}_{-1.39}$ & $76^{+30}_{-39}$ &  &  & $0.064^{+0.038}_{-0.040}$ & $0.051^{+0.030}_{-0.032}$ \\[5pt]
 intermediate & $11.30^{+0.79}_{-0.74}$ & $-0.59^{+1.75}_{-1.74}$ & $44^{+64}_{-54}$ &  &  & $0.033^{+0.039}_{-0.023}$ & $0.026^{+0.030}_{-0.018}$ \\[5pt]
 extreme & $10.82^{+0.71}_{-0.62}$ & $-4.56^{+1.19}_{-1.19}$ & $73^{+15}_{-15}$ &  &  & $0.154^{+0.043}_{-0.039}$ & $0.121^{+0.033}_{-0.030}$ \\
\hline
\end{tabular}

%% file: ngc6093_population_models.tex
\begin{tabular}{ccccccccccc}
\hline
population & $\Upsilon_{\rm 0}$ & $\Upsilon_{\rm t}$ & $\Upsilon_{\rm \infty}$ & $r_{\rm \Upsilon}$ & ${\overline q}$ & $\kappa_{\rm max}$ & $r_{\rm \kappa}$ & $\lambda_{\rm R,\,HL}$ & $\mu_{\rm \Delta_{C}}$ & $\sigma_{\rm \Delta_{C}}$ \\
& ${\rm M_\odot\,L_{\odot}^{-1}}$ & ${\rm M_\odot\,L_{\odot}^{-1}}$ & ${\rm M_\odot\,L_{\odot}^{-1}}$ & $\mathrm{arcmin}$ &  &  & $\mathrm{arcmin}$ &  &  & \\
\hline
 global & $4.0^{+1.7}_{-0.9}$ & $0.9^{+0.8}_{-0.5}$ & $2.4^{+1.6}_{-0.8}$ & $0.17^{+0.19}_{-0.13}$ & $0.83^{+0.05}_{-0.06}$ & $0.77^{+0.09}_{-0.09}$ & $0.77^{+0.40}_{-0.25}$ & $0.087^{+0.013}_{-0.010}$ &  &  \\[5pt]
 primordial &  &  &  &  &  & $0.68^{+0.23}_{-0.25}$ &  & $0.078^{+0.028}_{-0.030}$ & $0.063^{+0.003}_{-0.002}$ & $0.025^{+0.003}_{-0.002}$ \\[5pt]
 intermediate &  &  &  &  &  & $0.24^{+0.49}_{-0.52}$ &  & $0.043^{+0.056}_{-0.031}$ & $0.144^{+0.004}_{-0.004}$ & $0.016^{+0.005}_{-0.004}$ \\[5pt]
 extreme &  &  &  &  &  & $1.42^{+0.38}_{-0.44}$ &  & $0.145^{+0.031}_{-0.049}$ & $0.223^{+0.003}_{-0.003}$ & $0.027^{+0.003}_{-0.002}$ \\
\hline
\end{tabular}